\documentclass{PoS}

\usepackage{epstopdf}
\usepackage{cite}
\usepackage{amsmath,amssymb}

\newcommand{\be}{\begin{equation}}
\newcommand{\ee}{\end{equation}}
\newcommand{\ba}{\begin{eqnarray}}
\newcommand{\ea}{\end{eqnarray}}
\newcommand{\eq}{Eq.~}

\newcommand{\fig}{Fig.~}

\def\lsi{\raise0.3ex\hbox{$<$\kern-0.75em\raise-1.1ex\hbox{$\sim$}}}
\def\gsi{\raise0.3ex\hbox{$>$\kern-0.75em\raise-1.1ex\hbox{$\sim$}}}
\newcommand{\lsim}{\mathop{\lsi}}

\def\bfx{{\bf x}}

\def\tr{\mathop{\rm Tr}\nolimits}
\newcommand{\im}{{\rm Im}}

\title{
{\vspace{-25mm} \normalsize\hfill{\small CERN-PH-TH/2010-257}}\\[20mm]
Constraints for the QCD phase diagram from imaginary chemical potential}

\ShortTitle{Constraining the QCD phase diagram}

\author{\speaker{Owe Philipsen}\\%\thanks{A footnote may follow.}\\
        Institut f\"ur Theoretische Physik, Johann Wolfgang Goethe-Universit\"at Frankfurt,\\
        60438 Frankfurt am Main, Germany\\
        E-mail: \email{philipsen@th.physik.uni-frankfurt.de}}

\author{Philippe de Forcrand\\
        Institute for Theoretical Physics, ETH Z\"urich, CH-8093 Z\"urich, Switzerland\\
        Physics Department, TH Unit, CERN, CH-1211 Geneva 23, Switzerland\\
        E-mail: \email{forcrand@phys.ethz.ch}}

\abstract{We present unambiguous evidence from lattice simulations of $N_f=3$ QCD for
two tricritical points in the $(T,m)$ phase diagram at fixed imaginary $\mu/T=i\pi/3$ mod.~$2\pi/3$,
one in the light and one in the heavy quark regime. Together with similar results in the literature for $N_f=2$
this implies the existence of a chiral and of a deconfinement tricritical line at those 
values of imaginary chemical
potentials. These tricritical lines represent the boundaries of the analytically continued chiral and deconfinement critical
surfaces, respectively, which delimit the parameter space with first order phase transitions.
It is demonstrated that the shape of the deconfinement critical surface 
is dictated by tricritical scaling and implies the weakening of the deconfinement transition 
with real chemical potential. A qualitatively similar effect holds for the chiral critical surface.
}

\FullConference{The XXVIII International Symposium on Lattice Field Theory, Lattice2010\\
		June 14-19, 2010\\
		Villasimius, Italy}

\begin{document}

\section{Introduction}

\begin{figure}[t!!!]
\vspace*{-0.7cm}
\centerline{
\includegraphics[width=0.3\textwidth]{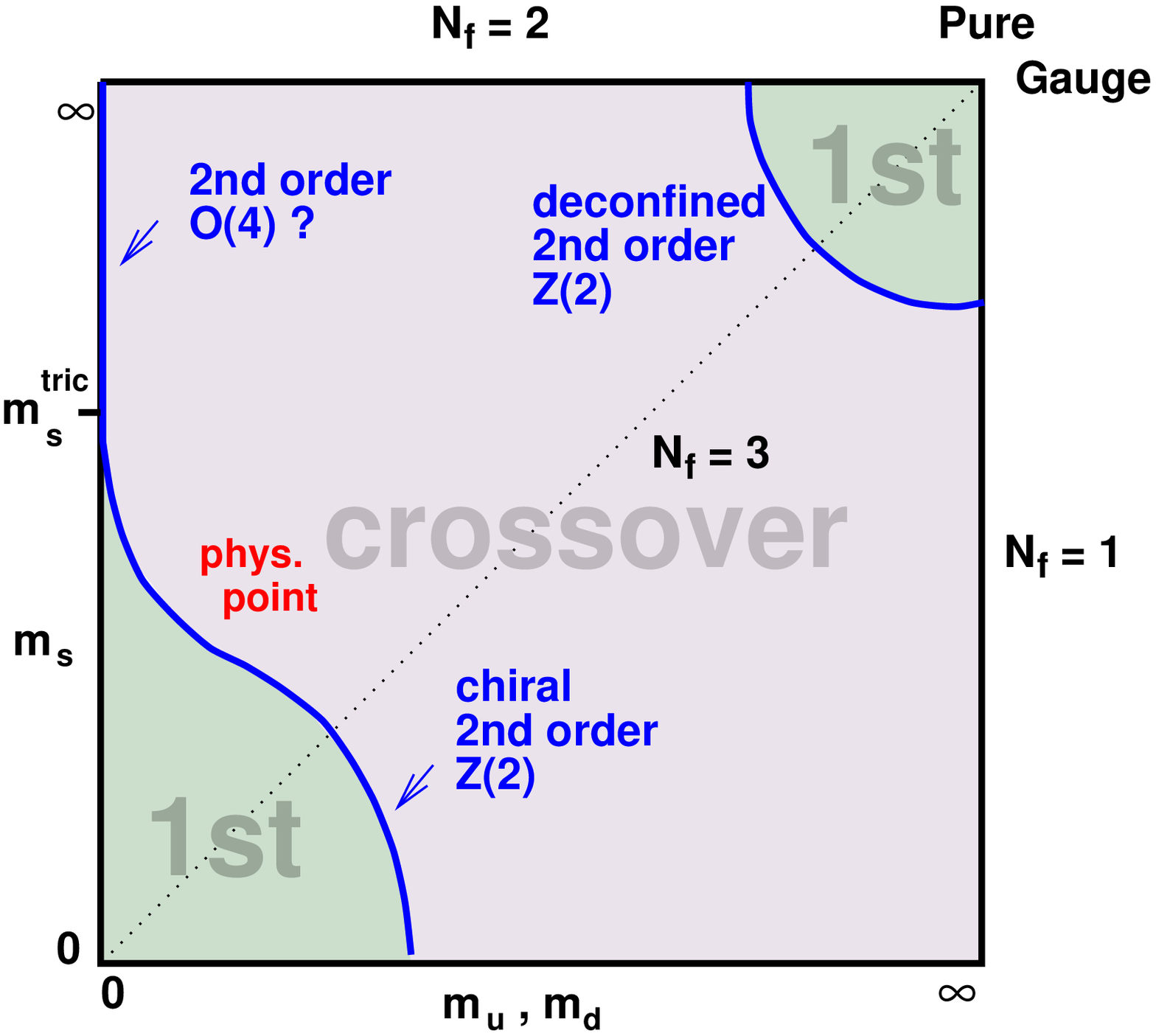}%\hspace*{1cm}
\includegraphics[width=0.42\textwidth]{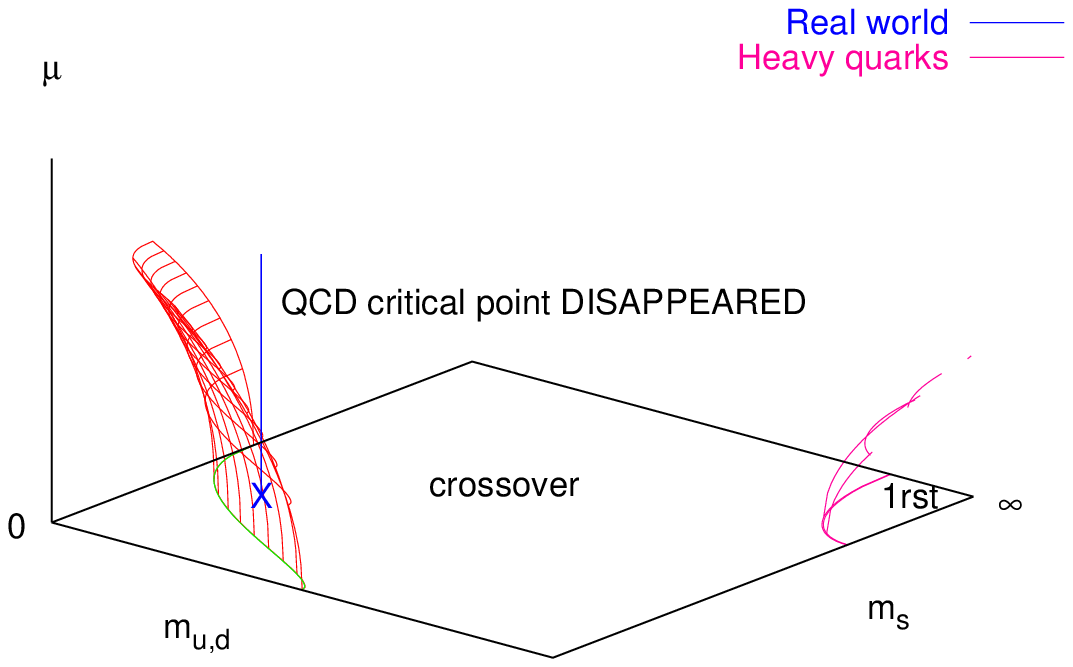}
\includegraphics[width=0.25\textwidth]{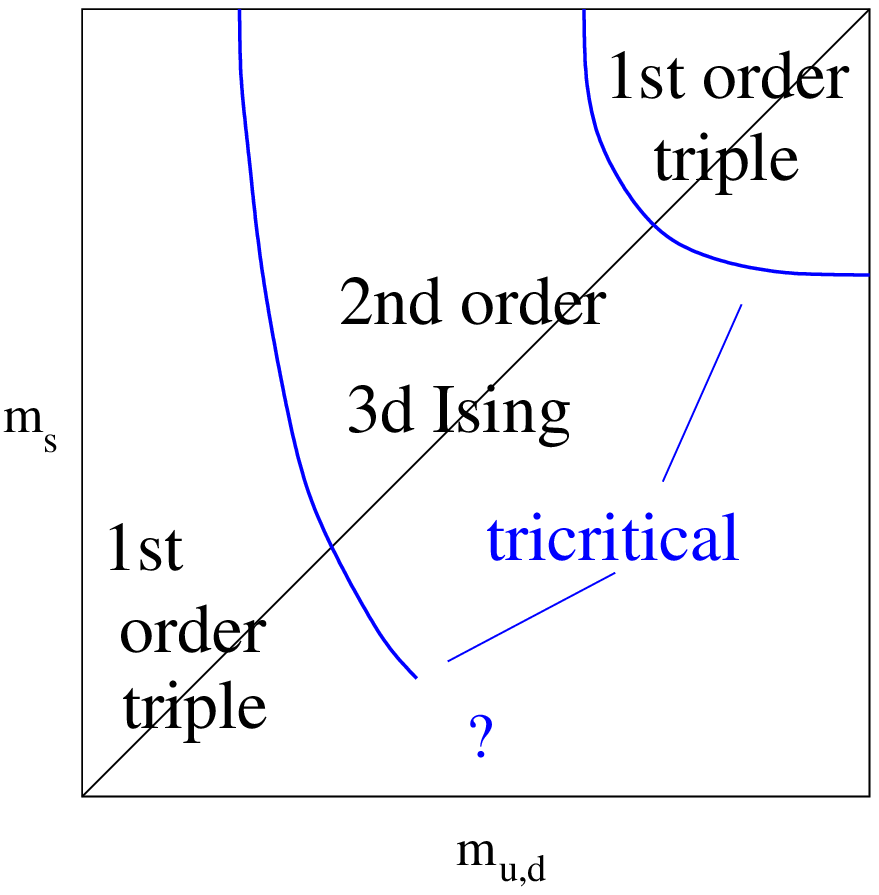}
}
%\vspace*{-3mm}
\caption[]{
Left: order of the quark hadron transition as a function of quark masses at $\mu=0$.
Middle: for finite $\mu$ the critical lines turn into
surfaces. The curvature is such that the chiral and deconfinement transitions are weakened.
Right: nature of the $Z(3)$-transition endpoint at $\mu/T=i\pi/3$.}
\label{col}
\end{figure}

Due to the sign problem prohibiting lattice simulations at finite baryon density, 
the QCD phase diagram in the space of temperature $T$ and chemical potential for baryon number
$\mu_B$ is largely unknown. 
Employing indirect methods like reweighting, Taylor expansions about
$\mu_B=0$ or simulations at imaginary chemical potentials $\mu=i\mu_i,\mu_i\in\mathbb{R}$, followed by 
analytic continuation, controlled calculations are only feasible as long as the quark chemical potential
$\mu=\mu_B/3\lsim T$ \cite{review}. 
Using the latter two methods we previously calculated the curvatures of the chiral
and deconfinement critical surfaces, 
which bound the mass regions that exhibit first order chiral or deconfinement  
transitions \cite{fp3,fp4,fkt,lp}. In both cases the curvature is such that the first order region shrinks, 
i.e.~the chiral and deconfinement phase transitions weaken with real chemical potential, as shown
schematically in \fig\ref{col} (left and middle).  

In this contribution we propose to study the phase diagram at imaginary chemical
potential, without continuing the numerical results directly to real $\mu$. Since 
the fermion determinant is real for imaginary chemical potentials, there is no sign problem
and simulations are feasible without additional systematic errors besides finite volume and 
cutoff effects, and at no additional computational cost compared to simulations at $\mu=0$.
For specific critical values of the imaginary chemical potential,  
there are rich critical structures like first order triple points, critical points with 3d Ising universality  
as well as tricritical points. 
We then argue that useful information for the phase diagram at real $\mu$ can be inferred from the
results.
In particular, we demonstrate that the weakening of the deconfinement transition in the heavy quark region is dictated by the tricritical scaling of the deconfinement critical surface at imaginary chemical potentials, 
with a similar weakening expected for 
the chiral transition.  

\section{The QCD phase diagram at imaginary chemical potential}

The QCD partition function exhibits
two important exact symmetries, reflection symmetry in $\mu$ and $Z(3)$-periodicity in $\mu_i$,
which hold for quarks of any mass \cite{rw},
\be
Z(\mu)=Z(-\mu), \quad 
Z\left(\frac{\mu}{T}\right) = Z\left(\frac{\mu}{T}+i\frac{2\pi n}{3}\right)\;,
\label{zsym}
\ee
for general complex values of $\mu$.
Let us now consider imaginary chemical potential, $\mu=i\mu_i$.
The symmetries imply transitions between adjacent centre sectors of the theory 
at fixed $\mu_i^c=(2n+1)\pi T/3, n=0,\pm 1,\pm 2,\ldots$.
The $Z(3)$-sectors are distinguished by the Polyakov loop 
\be
L(\bfx)=\frac{1}{3}{\tr} \prod_{\tau=1}^{N_\tau}U_0(\bfx,\tau)=|L|\,{\rm e}^{-i\varphi}\;,
\ee
whose phase $\varphi$ cycles through $\langle \varphi \rangle =n (2\pi/3), n=0,1,2,\ldots$
as the different sectors are traversed.
Moreover, the above also implies reflection symmetry about
the $Z(3)$ phase boundaries, $Z(\mu_i^c+\mu_i)=Z(\mu_i^c-\mu_i)$.

\begin{figure}[t]
%\hspace*{-0.5cm}
\vspace*{-0.5cm}
\centerline{
\includegraphics[width=0.3\textwidth]{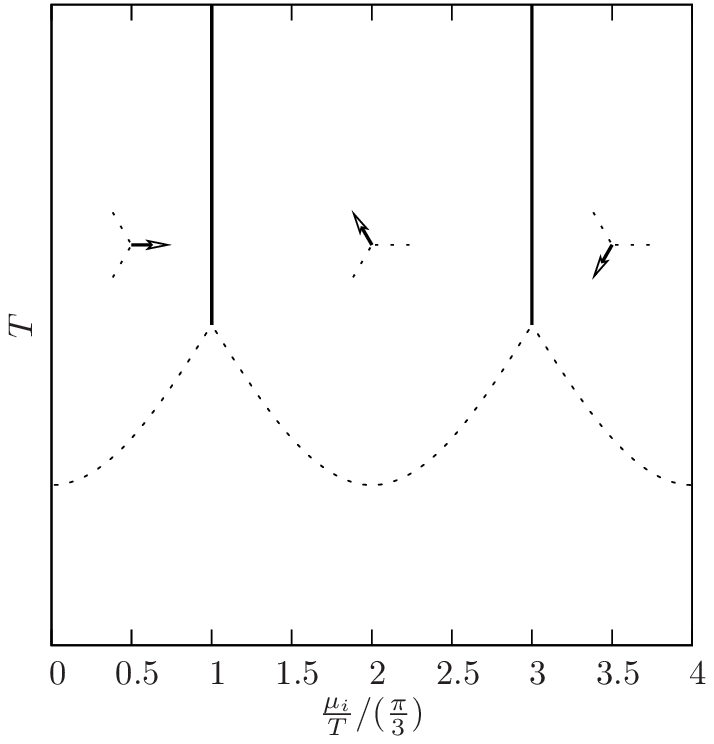}\hspace*{1.5cm}
\includegraphics[width=0.3\textwidth]{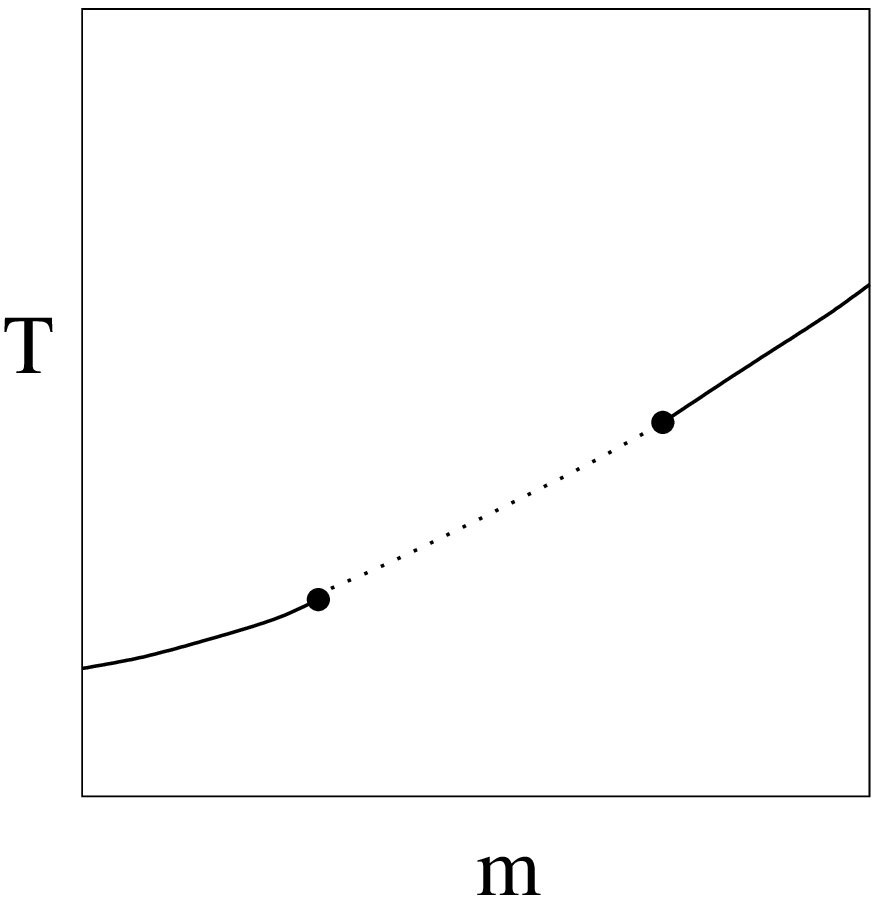}
\put(-85,80){$\langle \phi \rangle\neq 0$}
\put(-55,25){$\langle \phi \rangle =0$}
}
%\vspace*{-0.5cm}
\caption[]{Left: phase diagram for imaginary $\mu$. Vertical lines are
first order transitions between $Z(3)$-sectors, arrows show the phase
of the Polyakov loop. The $\mu\!=\!0$ chiral/deconfinement transition continues to imaginary $\mu$,
its order depends on $N_f$ and the quark masses.
Right: phase diagram for $N_f=3$ at $\mu=i\pi T$. Solid lines are lines of triple points
ending in tricritical points,  connected by a $Z(2)$ critical line.}
\label{schem}
\end{figure}

Transitions in $\mu_i$ between neighbouring sectors 
are of first order for high $T$ and analytic crossovers
for low $T$ \cite{rw,fp1,el1}, as shown in \fig\ref{schem} (left).  
Correspondingly, for fixed $\mu_i=\mu_i^c$, there are transitions in $T$
between an ordered phase with two-state coexistence at high $T$ and a disordered phase at low $T$.
An order parameter to distinguish these phases is the
shifted phase of the Polyakov loop, $\phi=\varphi-\mu_i/T$ \cite{yah}. 
At high temperature it fluctuates 
about $\langle \phi \rangle=\pm \pi/3$ 
on the respective sides of $\mu_i^c$. The thermodynamic limit picks one of those states,
thus spontaneously breaking the reflection symmetry 
about $\mu_i^c$. 
At low temperatures $\phi$ fluctuates smoothly between those values, with the 
symmetric ground state $\langle \phi\rangle =0$.

Away from $\mu_i=\mu_i^c$, there is a chiral or deconfinement
transition line separating high and low temperature regions. This line represents the analytic
continuation of the chiral or deconfinement  transition at real $\mu$. Its nature 
(1st, 2nd order or crossover) depends on the number of quark flavours and masses.
It has long been believed that this line meets the $Z(3)$ transition at its endpoint, and  
early evidence \cite{fp1,el1} is consistent with this.
While a lot of numerical work at imaginary chemical potential was devoted to determining
the chiral or deconfinement transition and continue it about $\mu=0$, here we are interested
in the nature of the endpoint of the $Z(3)$ transition line as a function of quark masses.
Similar investigations have been carried out for $N_f=4$ \cite{el2}
and more recently for $N_f=2$ \cite{mass}.  We thus fix the chemical potential to 
an imaginary critical value, $\mu_i=\pi T$, and investigate the order of the transition by scanning
vertically in $T$ for various masses.

\section{Numerical results for $N_f=3$}

In this investigation we consider $N_f=3$ QCD, using the standard staggered action and the 
RHMC algorithm. In order to identify the order of the transition, we study the 
finite size scaling
of the Binder cumulant constructed from the imaginary part of the Polyakov loop, 
\be 
B_4(\im (L)) \equiv \langle [\im (L) - \langle \im (L) \rangle]^4 \rangle 
/ \langle [\im (L) - \langle \im (L) \rangle]^2 \rangle^2
= \langle (\im (L))^4 \rangle / \langle (\im (L))^2 \rangle^2\;.
\ee
For $\mu/T\!=\!i\pi$, every $\beta$-value represents a point on the phase boundary and thus
is pseudo-critical. In the thermodynamic limit, $B_4(\beta)\!=\!3, 1.5, 1.604, 2$ for crossover, 
first order triple point, 3d Ising and tricritical transitions, respectively. 
On finite $L^3$ volumes the steps between these
values are smeared out to continuous functions whose gradients increase with volume.
The critical coupling $\beta_c$ for the endpoint is obtained as the intersection of curves from
different volumes. In the scaling region around $\beta_c$,  
$B_4$ is a function of $x=(\beta-\beta_c)L^{1/\nu}$ alone and can be expanded 
\be
B_4(\beta,L)=B_4(\beta_c,\infty)+a_1 x + a_2 x^2+\ldots,
\label{scale}
\ee
up to corrections to scaling,
with the critical exponent $\nu$ characterising the approach to the thermodynamic limit.
The relevant values for us are $\nu=1/3, 0.63, 1/2$ 
for a first order, 3d Ising or tricritical transition, respectively.
\begin{figure}
\vspace*{-0.7cm}
\centerline{
\includegraphics[width=0.45\textwidth]{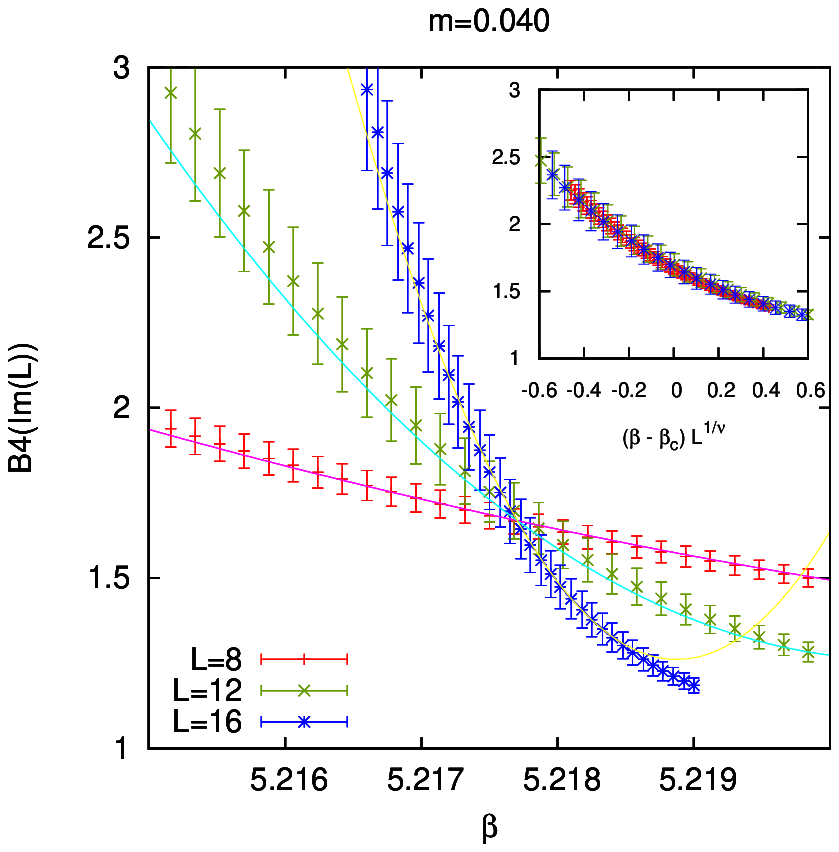}
\includegraphics[width=0.45\textwidth]{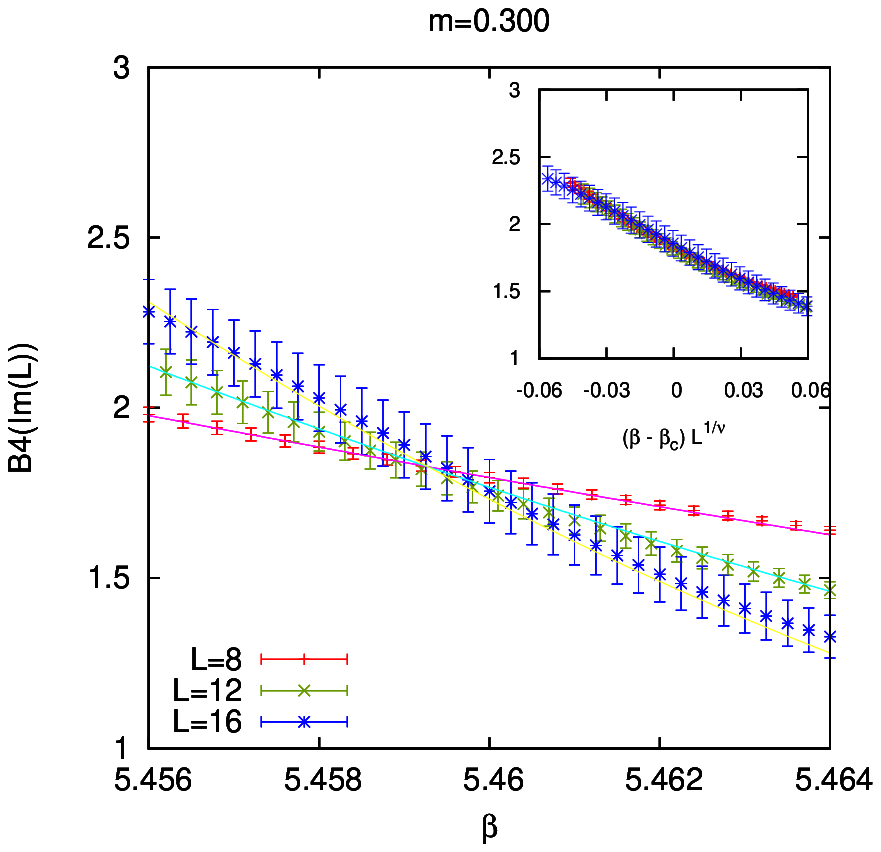}
}
\caption[]{Finite size scaling of $B_4$ for small and intermediate quark masses, fitted to \eq(\ref{scale}). Insets
show data rescaled with fixed $\nu=0.33, 0.63$, 
corresponding to a first/second order transition, respectively.}
\label{b4}
\end{figure}

For each quark mass, we simulated lattices of sizes $L=8, 12, 16$ (20 in a few cases), at typically 8-14 different $\beta$-values,
calculated $B_4({\rm Im}(L))$ and 
filled in additional points by Ferrenberg-Swendsen reweighting \cite{fs}.
\fig\ref{b4} shows examples for quark masses $am\!=\!0.04, 0.3$.
$B_4$ moves from large values (crossover) at small $\beta$ (i.e.~low $T$)
towards 1 (first order transition) at large $\beta$ (i.e.~high $T$). 
In the neighbourhood of the intersection 
point, we then fit all curves simultaneously to
\eq(\ref{scale}), thus extracting $\beta_c,B_4(\beta_c,\infty),\nu, a_1, a_2$.
We observe that the value of the Binder cumulant at the intersection can be far from the expected
universal values in the thermodynamic limit. This is a common situation:
large finite-size corrections are observed in simpler spin models even
when the transition is strongly first-order \cite{1st}. Moreover, in our case,
logarithmic scaling corrections will occur near a tricritical point since
$d=3$ is the upper critical dimension in this case \cite{ls}.
Fortunately, the critical exponent $\nu$, which determines the 
approach to the thermodynamic limit, is less sensitive to
finite-size corrections and in \fig\ref{b4} consistent with $\nu=0.33, 0.63$, its values for first and
second order transitions, respectively. A check is to fix $\nu$ to one of the universal values 
and see whether the curves collapse under the appropriate rescaling, as in \fig\ref{b4} insets. 
Note that the critical coupling determined from the intersection of the
$B_4$ curves in \fig\ref{b4} is consistent with the one extracted from the
peak of the specific heat or the chiral susceptibility.

\begin{figure}[t]
\vspace*{-0.5cm}
%\hspace*{-0.9cm}
\includegraphics[width=0.45\textwidth]{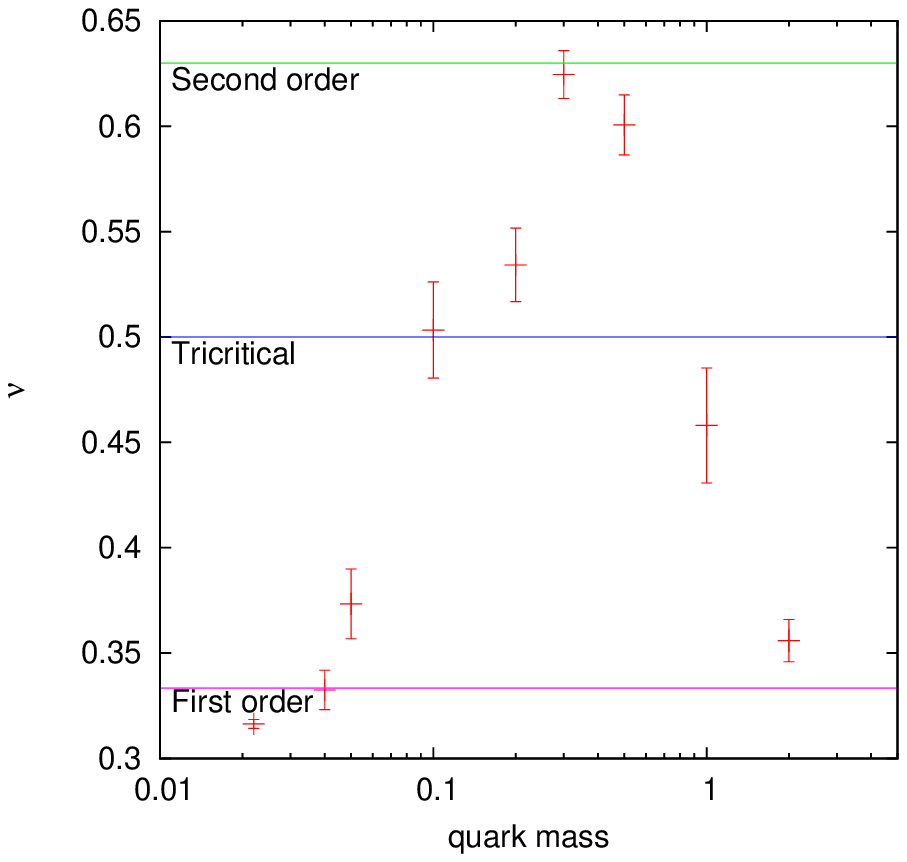}%\hspace*{-1.2cm}
\includegraphics[width=0.45\textwidth]{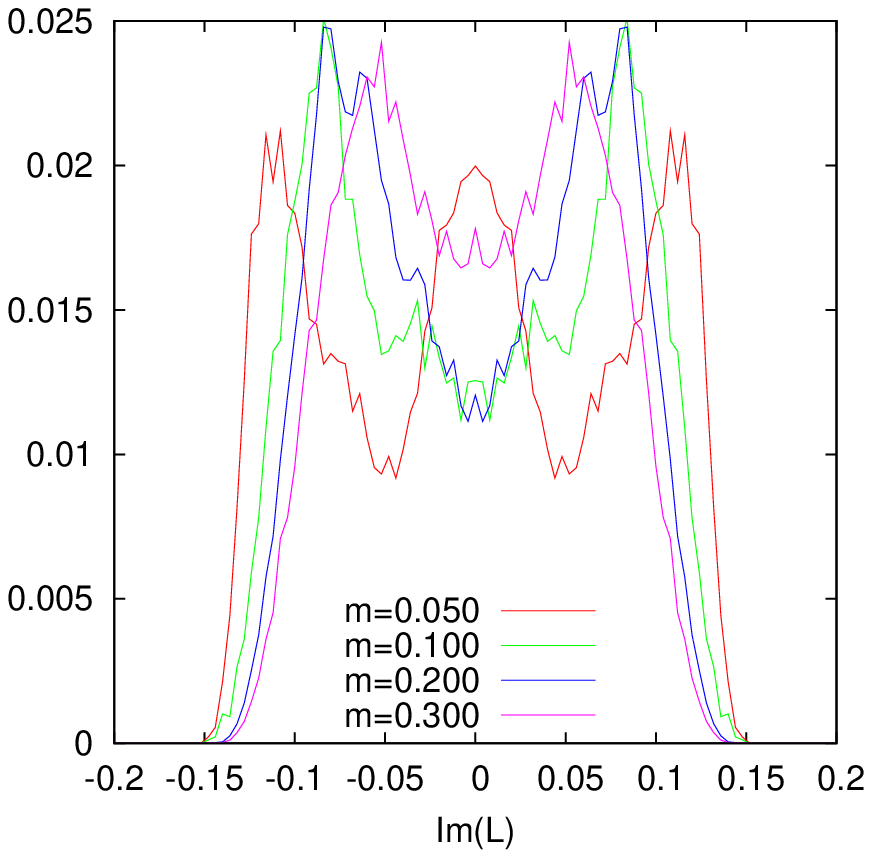}
\caption[]{Left: Critical exponent $\nu$ at $\mu/T=i\pi$. Right: ${\rm Im}(L)$ distribution at the
 $Z(3)$ transition endpoint.}
\label{exp}
\end{figure}

We have investigated quark mass values ranging from the chiral to the pure gauge
regime. The exponents $\nu$ pertaining to each of them are shown in  \fig\ref{exp} (left).
There is unambiguous evidence for a change from first order scaling to 3d Ising scaling, 
and back to first order scaling.  
Note that, in the infinite volume limit, the curve would be replaced by a non-analytic step function, whereas the smoothed-out rise and fall in \fig\ref{exp} (left) corresponds to finite volume corrections.

The results from the finite size scaling of $B_4$ can be sharpened by 
the probability distribution of ${\rm Im}(L)$ at the critical couplings $\beta_c$, 
corresponding to the crossing points. This is shown in \fig\ref{exp} (right) for 
masses $am=0.05,0.1,0.2,0.3$ for $L=16$.
The lightest mass  
displays a clear three-peak structure, indicating
coexistence of three states 
at the coupling 
$\beta_c$, which therefore corresponds to a triple point. The same observation holds for 
heavy masses. For $am=0.1,0.2$ the central peak is disappearing and 
for $am=0.3$ we are left with the two peaks characteristic for the magnetic direction 
of 3d Ising universality.  
We have checked the expected volume-scaling of all distributions.

Hence, for small and large masses, we have unambiguous evidence that
the boundary point between a first order $Z(3)$ transition and a crossover at $\mu=i\pi T$ corresponds to a triple point. 
This implies that two additional first order lines branch off the $Z(3)$-transition line as in 
\fig\ref{schem} (left), which are to be identified as the chiral (for light quarks) or deconfinement (for heavy 
quarks) transition at imaginary chemical potential. 
This is expected on theoretical grounds: for $m=0$ or $+\infty$, these 
transitions are first-order for any chemical potential.
The fact that the endpoint of the $Z(3)$ transition line changes its nature from a triple point at 
low and high masses to 
second order for intermediate masses implies the existence of two tricritical points. 

We are thus ready to discuss the $(T,m)$ phase diagram of $N_f=3$ QCD at fixed imaginary chemical
potential, $\mu=i(2n+1)\pi T/3$. The qualitative situation is shown in \fig\ref{schem} (right).
For high temperatures, we have a two-phase coexistence with the phase of the Polyakov loop
flipping between two possible values.  At low temperatures, instead, we observe phase
averaging over the possible phases of the Polyakov loop.
Since the transition between these regimes is associated with a breaking of a global
symmetry, it is always non-analytic. 

An important question concerns 
cut-off effects. These strongly affect quark masses, and in particular
the tricritical  points. 
However, universality implies that critical
behaviour is insensitive to the cut-off, as long as the global symmetries 
of the theory are not changed. 
Our calculation is therefore 
sufficient to establish the qualitative picture \fig\ref{schem} (right) in the continuum.
The change from first order to 3d Ising behaviour for low and intermediate masses has been
observed earlier for $N_f=2$ \cite{mass} and we expect the corresponding $(T,m)$-diagram to look
the same.
\begin{figure}[t!!!]
\vspace*{-0.5cm}
\hspace*{-0.9cm}
\includegraphics[width=0.34\textwidth]{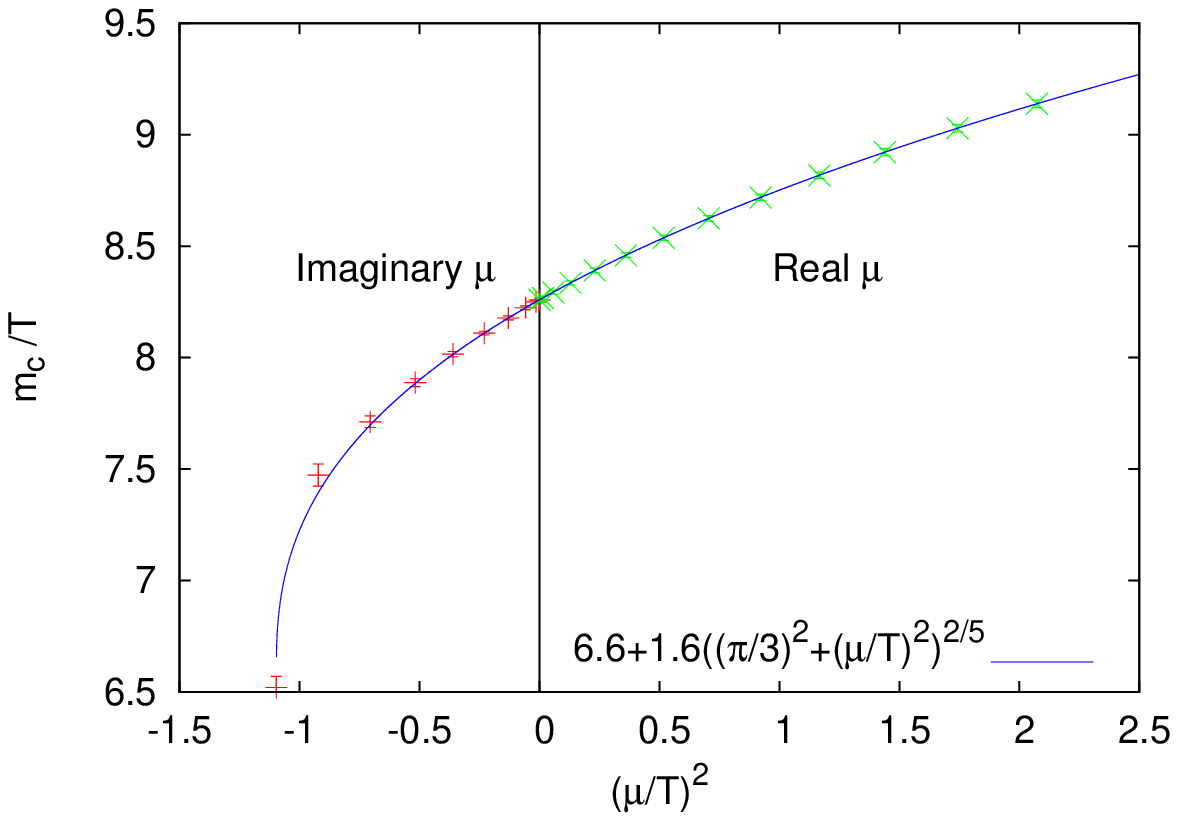}
\includegraphics[width=0.34\textwidth]{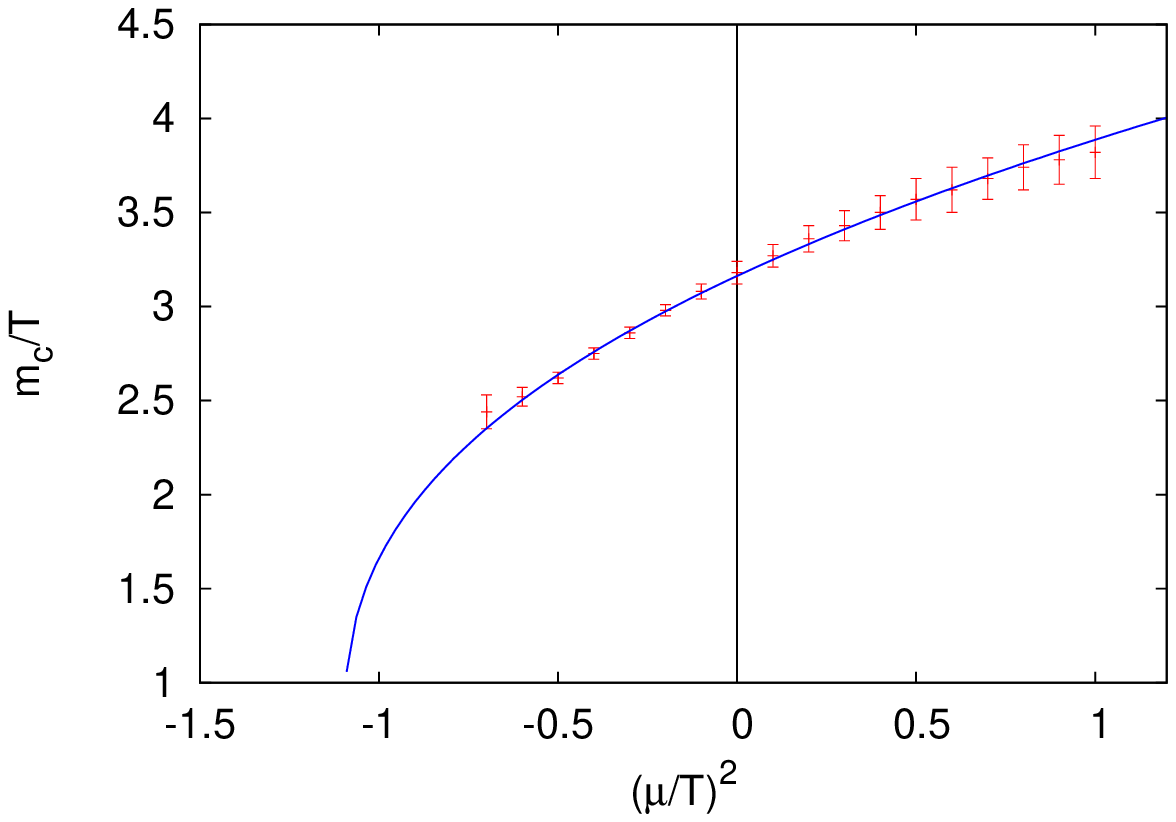}
\includegraphics[width=0.34\textwidth]{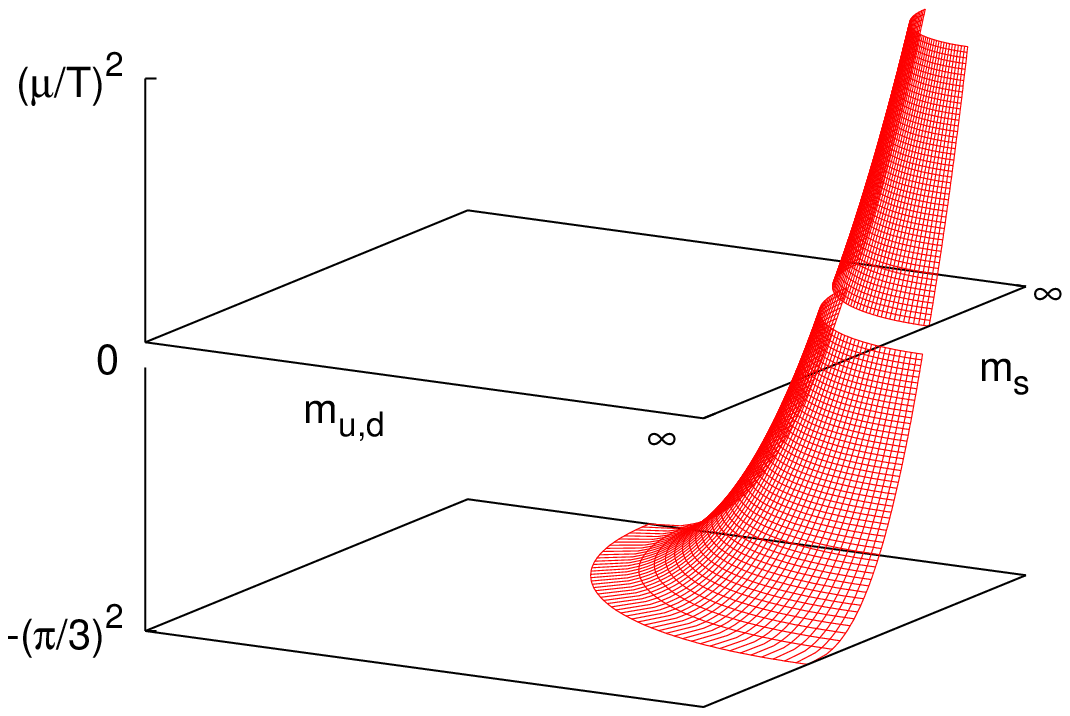}
%\vspace*{-0.4cm}
\caption[]{Critical line $m_c(\mu^2)$ in the 3-state Potts model~\cite{fkt} (left) and
for QCD in a strong 
coupling expansion~\cite{lp} (middle). Right: deconfinement critical surface determined by 
tricritical scaling.}
\label{tric}
\end{figure}

\section{$N_f=2+1$ and connection to real $\mu$}

Combining our knowledge of $N_f=2,3$, the nature of the 
$Z(3)$ transition endpoint can be characterised as a function of quark masses as in \fig\ref{col} (right), 
in complete analogy to the 
corresponding plot at $\mu=0$ (left). Schematically, we have a first order region of triple
points for both heavy and light masses, which are separated from a region of second order points
by a chiral and deconfinement tricritical line, respectively. This entire diagram
is computable by standard Monte Carlo methods and constitutes a useful
benchmark for model studies of the QCD phase diagram.

How is this diagram connected to the one at $\mu=0$?
Generally, a tricritical point
represents the confluence of two ordinary critical points. In the heavy mass region 
the critical endpoints of the deconfinement transition, representing the 
deconfinement critical surface, merge with the endpoints of the $Z(3)$ transition. 
Thus, the deconfinement tricritical line is the boundary of the deconfinement critical
surface at $\mu=i\pi T/3$. This can be explicitly demonstrated by simulations of the 3d  three-state
Potts model. It is well know that this model is in the same universality class as QCD with heavy quarks
and can therefore be used in the neighbourhood of the deconfinement critical line. In particular,
it has been used to calculate, for a fixed number of flavours, the change of the critical mass
with chemical potential, since the sign problem is mild and manageable there \cite{fkt}.
The results, including a tricritical point, are shown in \fig\ref{tric} (left), together with a QCD strong coupling
expansion result (right) \cite{lp}. 
The deviation from the symmetry plane, $((\mu/T)^2+(\pi/3)^2)$,
is analogous to an external field in a spin model, and the way a critical line leaves a tricritical point
in such a field is again universal \cite{ls},
\be
\frac{m_c}{T}(\mu^2)=\frac{m_{\rm tric}}{T} + K \left[\left(\frac{\pi}{3}\right)^2+\left(\frac{\mu}{T}\right)^2\right]^{2/5}\;.
\label{mean}
\ee
\fig\ref{tric} shows that the data from \cite{fkt,lp} excellently fit this form, 
far into the real chemical potential region. 
Thus for heavy quark masses, the form of the critical
surface of the deconfinement transition is  determined by tricritical scaling of the $Z(3)$ transition at imaginary $\mu=i\pi T/3$.

It is clear that the chiral critical surface will likewise 
terminate on the chiral tricritical line at $\mu=i\pi T/3$.
Unfortunately, for this surface no suitable effective model  
is available and we presently do not know
to which extent it is shaped by tricritical scaling. 
Estimating $am_{\rm tric1}\sim 0.1$ and using $am_c(0)\approx 0.0265$ \cite{fp4}, 
$K$ is fixed and expansion
of \eq(\ref{mean}) predicts a negative curvature $c_1\approx -10$ for the chiral critical surface, 
as compared to the directly calculated $c_1=-3.3(3)$ (in the notation of \cite{fp4}).
Tricritical scaling thus predicts a weakening also of the chiral phase transition with real chemical 
potential, independently confirming the findings in \cite{fp1,fp3,fp4}.

\section{Conclusions}

We have clarified the nature of the endpoint of the Roberge-Weiss or $Z(N)$ transition at imaginary
chemical potentials as a function of quark masses and firmly established that it connects with the 
(pseudo-) critical lines of the chiral or deconfinement transition. For light and heavy quark masses,
the latter are of first order and the junction is a triple point, while it is a critical endpoint in the 3d 
universality class otherwise. We have generalised this result to arbitrary quark mass combinations
and sketched a ``Columbia plot'' for $\mu=i\pi T/3$. The plot features two tricritical lines bounding
areas of triple points, which represent the boundaries of the chiral and deconfinement critical surfaces,
respectively. We further demonstrated that the curvature of the deconfinement critical surface
is determined by the associated tricritical scaling and argued the same to hold qualitatively for the negative curvature of the chiral critical surface.\\

\noindent
{\bf Acknowledgement:}  This project is partially supported by the German BMBF, 06MS9150. 
We thank the University of Minnesota Supercomputer Institute for computing resources.

\end{document}